\documentclass{article}

\usepackage[dblblindworkshop, final, nonatbib]{neurips_2025_libribrain_competition}

\workshoptitle{2025 PNPL Competition}

\usepackage[utf8]{inputenc} 
\usepackage[T1]{fontenc}    
\usepackage{hyperref}       
\usepackage{url}            
\usepackage{booktabs}       
\usepackage{amsfonts}       
\usepackage{nicefrac}       
\usepackage{microtype}      
\usepackage{xcolor}         
\usepackage{graphicx}
\usepackage{amsmath}

\title{MEBM-Speech: Multi-scale Enhanced BrainMagic for Robust MEG Speech Detection}

\author{%
  Li Songyi$^{1}$ \And Zheng Linze$^{1}$ \And Liang Jinghua$^{1}$ \And Zhang Zifeng$^{1,2}$
  \\
  $^{1}$Speech and Hearing Research Center, School of Intelligence Science and Technology \\
  $^{2}$Center for BioMed-X Research, Academy for Advanced Interdisciplinary Studies \\ 
  Peking University, Beijing, China 100871\\
  \texttt{2501213415@stu.pku.edu.cn} \\
}

\begin{document}

\maketitle

\begin{abstract}
We propose \textbf{MEBM-Speech}, a multi-scale enhanced neural decoder for speech activity detection from non-invasive magnetoencephalography (MEG) signals.
Built upon the BrainMagic backbone, MEBM-Speech integrates three complementary temporal modeling mechanisms: a multi-scale convolutional module for short-term pattern extraction, a bidirectional LSTM (BiLSTM) for long-range context modeling, and a depthwise separable convolutional layer for efficient cross-scale feature fusion.
A lightweight temporal jittering strategy and average pooling further improve onset robustness and boundary stability.
The model performs continuous probabilistic decoding of MEG signals, enabling fine-grained detection of speech versus silence states—an ability crucial for both cognitive neuroscience and clinical applications.
Comprehensive evaluations on the \textbf{LibriBrain Competition 2025 Track~1} benchmark demonstrate strong performance, achieving an average F1\textsubscript{macro} of 89.3\% on the validation set and comparable results on the official test leaderboard.
These findings highlight the effectiveness of multi-scale temporal representation learning for robust MEG-based speech decoding.
\end{abstract}

\section{Introduction}
\label{instruction}
Understanding how the human brain encodes and processes acoustic and linguistic information underlying speech perception is a long-standing goal in cognitive neuroscience and brain–computer interface (BCI) research.  
Recent advances in magnetoencephalography (MEG) decoding open new possibilities for non-invasively mapping brain dynamics to acoustic and linguistic features \cite{moses2016, defossez2023}.  
Accurately decoding speech-related neural activity not only deepens our understanding of language processing but also holds important clinical implications.  
Real-time detection of speech and silence states from MEG can enable monitoring of residual speech perception in patients with neurological impairments and facilitate detection of speech intention in locked-in or aphasic individuals, contributing to communication-restoring BCIs \cite{dash2020}.  

The LibriBrain Competition 2025 \cite{landau2025,ozdogan2025} provides a large-scale benchmark for decoding acoustic and linguistic features from non-invasive brain recordings.  
In particular, Track~1 focuses on detecting speech versus silence segments from MEG data while participants listen to natural audiobook stimuli.  
This task requires the model to learn the complex temporal dynamics underlying speech and silence auditory processing and generalize effectively across speakers and contexts.

To address this task, we propose a novel model termed MEBM-Speech, specifically designed for Track~1 of the LibriBrain Competition.  

Our approach introduces three key innovations:
\begin{enumerate}
    \item \textbf{Decoding Strategy:}
    Unlike the official baseline method that formulates the task as framewise binary classification, our model performs end-to-end probabilistic decoding, predicting a continuous probability for each frame within a temporal window, followed by adaptive thresholding to determine speech versus silence regions.
    A similar continuous decoding perspective has proven effective in recent EEG-to-audio reconstruction work~\cite{xu2024}, which motivates our design choice for temporally coherent probabilistic modeling.
    
    \item \textbf{Model Architecture:} We augment the BrainMagic~\cite{defossez2023} backbone with a short-term multi-scale convolutional module for capturing fine-grained temporal features, a BiLSTM layer for modeling long-term dependencies, and a depthwise separable convolution layer for efficient feature fusion.  

    \item \textbf{Training Protocol:} We adopt a 100\,Hz downsampled MEG signal representation and use only the gradient (\texttt{grad}) channels for prediction.  
    This design significantly reduces computational and memory costs while preserving critical spatial-temporal information, leading to faster convergence without degrading performance.
\end{enumerate}

\section{Methods}
\label{methods}
\subsection{Decoding Strategy}
\label{sec:decoding}
In the speech detection task, the auditory stimuli alternate continuously between speech and silence states, and the corresponding MEG signals dynamically adjust to these changing acoustic states.
Instead of treating each frame as an independent binary classification target, we formulate the task as a continuous probabilistic decoding problem,
where the model predicts a time-varying probability sequence $\mathbf{P} \in [0,1]^{1\times T}$ representing the likelihood of speech activity at each time point.

This design better aligns with the intrinsic continuity of brain activity:  
neural responses to speech do not abruptly switch between binary states but evolve smoothly in time with overlapping temporal integration windows.  
By learning continuous probabilities, the model can capture gradual onset and offset transitions, improving temporal precision and robustness to boundary ambiguity.  
During inference, an adaptive thresholding scheme is applied to $\mathbf{P}$ to obtain final binary predictions for speech versus silence segments.  

To further enhance temporal robustness, we apply a mild temporal jittering when generating training labels.
For each phoneme onset, the starting point is randomly shifted within the range $[\text{onset}-2, \text{onset}+2]$ frames (corresponding to $\pm 20$\,ms at 100\,Hz),
ensuring the model remains invariant to small onset misalignments inherent in MEG signals.

During training, the model is optimized using a mean squared error (MSE) loss between the predicted probability sequence 
and the ground-truth binary speech labels. 
For model selection, the five checkpoints with the lowest validation losses are retained.  
Each candidate model is then evaluated on the validation set using 99 classification thresholds ranging from 0.01 to 0.99 
(with a step size of 0.01).  
The model-threshold combination achieving the highest F1\textsubscript{macro} score is selected as the final configuration for subsequent testing.

\subsection{Model Architecture}
\label{sec:model}

\begin{figure}[t]
  \centering
  \includegraphics[width=0.9\linewidth]{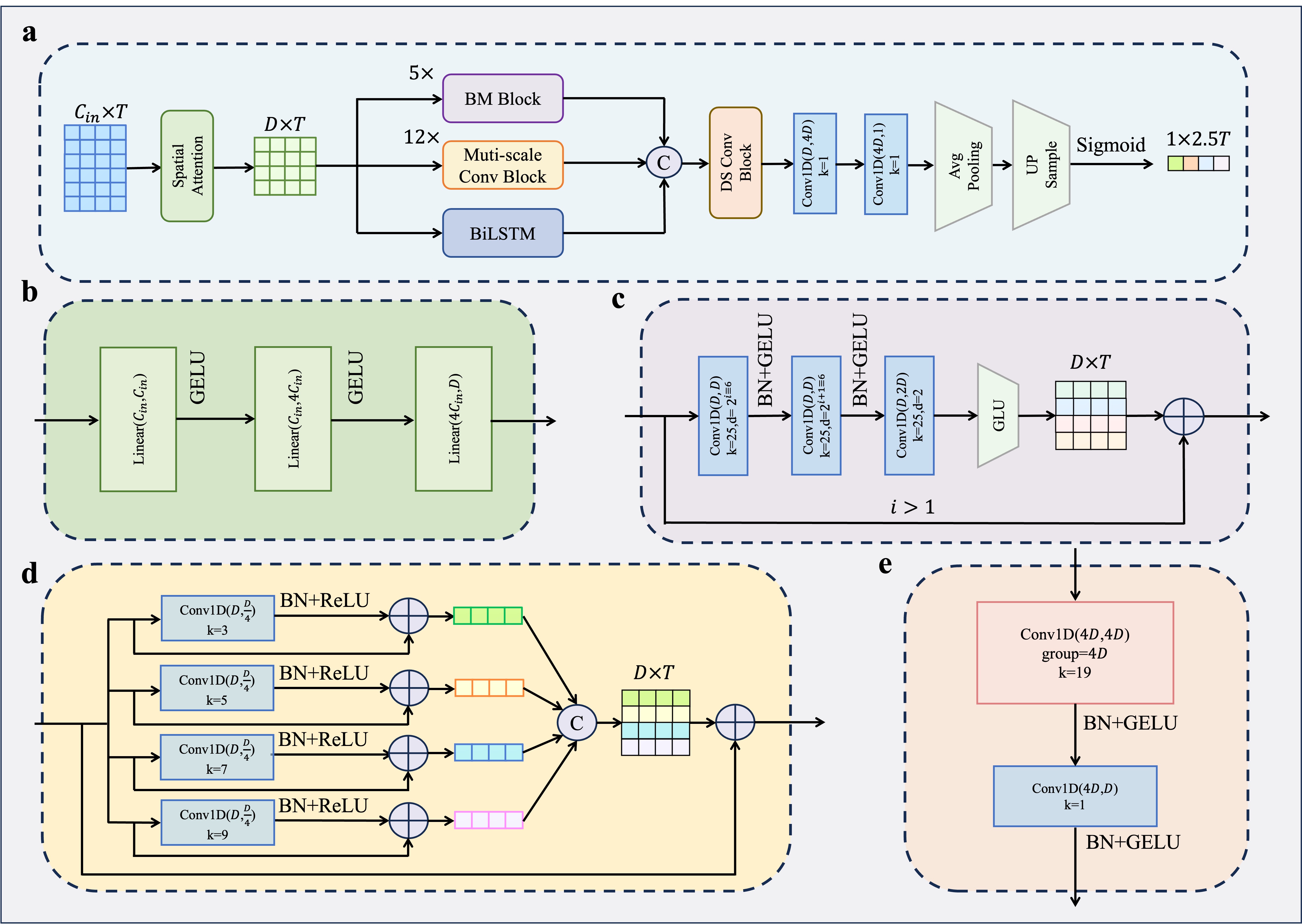}
  \caption{
  Overall architecture of the proposed MEBM-Speech model.  
  \textbf{(a)} The complete processing pipeline.
  \textbf{(b)} The \textit{spatial attention module} enhances sensor-level representations by learning spatial relevance weights across MEG channels.  
  \textbf{(c)} The \textit{BM encoder} extracts mid-term contextual features from spatially weighted signals.  
  \textbf{(d)} The \textit{short-term multi-scale convolutional module} captures fine-grained temporal dependencies using multiple receptive fields.  
  \textbf{(e)} The \textit{depthwise separable convolutional layer} further refines temporal representations with lightweight channel-wise and pointwise filtering.
  }
  \label{fig:model}
\end{figure}

As illustrated in Figure~\ref{fig:model}, the proposed MEBM-Speech architecture extends the BrainMagic framework by incorporating enhanced short-term temporal modeling and multi-scale feature integration.
Given the MEG input $\mathbf{X} \in \mathbb{R}^{C_{\text{in}} \times T}$, where $C_{\text{in}}$ denotes the number of MEG sensor channels, and $T$ represents the total number of temporal samples, the model first passes through a spatial attention module composed of multiple linear and activation layers, which dynamically recalibrates channel responses to produce a spatially refined representation $\mathbf{H}_s \in \mathbb{R}^{D \times T}$, with $D$ being the dimensionality of the projected feature space.
The refined representation is then processed in parallel by three temporal branches:
(1) 5 BrainMagic (BM) encoders that capture mid-term contextual dependencies,
(2) 12 multi-scale convolutional blocks that extract fine-grained local features across multiple receptive fields, and
(3) a BiLSTM that models long-range temporal dependencies.
The outputs of these three branches are concatenated along the feature dimension to form a unified temporal representation encompassing multi-scale contextual information.
A depthwise separable convolutional block is then applied to efficiently fuse and refine the concatenated features while reducing inter-channel redundancy.
Subsequent average pooling provides temporal smoothing and stabilizes boundary estimation.
Since the MEG data are downsampled to 100\,Hz, the final probabilistic predictions are temporally upsampled via linear interpolation, yielding a continuous per-sample probability of speech activity.
This hierarchical fusion framework enables MEBM-Speech to capture transient neural responses and sustained temporal dependencies simultaneously, thereby enhancing the model’s ability to detect speech-related neural activity in continuous MEG recordings with high temporal fidelity and robustness.

\section{Experiments}
\label{experiments}
\subsection{Experimental Setup}
\label{sec:setup}
The offline validation set was constructed using the official validation and test sessions (\textit{Sherlock1, sessions 11–12}) to approximate the holdout distribution defined by the LibriBrain challenge.  
For consistency, both the training and validation data followed the same preprocessing pipeline.  
The continuous MEG signals of each session were first downsampled to 100\,Hz and normalized independently along the temporal dimension.  
Only the \textit{grad} channels were retained for subsequent processing, resulting in the input dimensionality of $C_{\text{in}} = 204$.  
For each sample, MEG segments were extracted using a 12-second window ($T = 1200$) with a 6-second step size, yielding overlapping windows to augment the effective training data volume.
Each 12-second segment was normalized separately to maintain session-level stability.

The proposed MEBM-Speech model was implemented in PyTorch and trained on a single NVIDIA A800 GPU (80\,GB).  
The network comprised approximately 10.3\,M trainable parameters and converged within 10 epochs, requiring roughly 20 minutes of training.  
The intermediate feature dimension was set to $D = 128$ with a dropout rate of 0.01.  
All convolutional operations employed \texttt{padding='same'} to preserve temporal resolution. The average pooling layer employed a window size of 31 and a stride of 15.
Training was performed using the AdamW optimizer with a learning rate of $1\times10^{-3}$.  
Model selection and hyperparameter tuning were conducted using the offline validation set derived from the \textit{Sherlock1 Session 11–12} data.

\subsection{Results and Ablation}
\label{sec:results}

We report the performance of the proposed MEBM-Speech model and its ablated variants on the offline validation set.
All results are averaged over six random seeds ${0,1,2,3,4,5}$ for reproducibility.
Evaluation metrics include F1\textsubscript{macro} (\%) and Acc\textsubscript{macro} (\%).
Table~\ref{tab:ablation} summarizes the averaged results across all seeds.
The full MEBM-Speech model achieves an average F1\textsubscript{macro} of 89.34\% and Acc\textsubscript{macro} of 89.25\% on the validation set.
When submitted to the official online test server, it attains a comparable performance of approximately 89\% F1\textsubscript{macro}, demonstrating strong generalization to unseen sessions.

Removing the BM Encoder leads to the largest performance drop, confirming its central role in capturing mid-term spatiotemporal representations and encoding discriminative brain dynamics.
Excluding the Multi-scale Convolution branch or the BiLSTM results in moderate yet consistent degradation, suggesting that both modules contribute complementary temporal features—short-term fine-grained cues from convolutional filters and long-range contextual dependencies from recurrent modeling.
The fusion of these components within the full architecture yields the most balanced and robust decoding performance across all evaluation metrics.

\begin{table}[t]
\centering
\caption{Results and ablation analysis on the local validation set under six random seeds (0–5). 
Metrics include F1\textsubscript{macro} and Acc\textsubscript{macro} (mean ± std).}
\label{tab:ablation}
\begin{tabular}{lcc}
\toprule
\textbf{Model Variant} & \textbf{F1\textsubscript{macro} (\%)} & \textbf{Acc\textsubscript{macro} (\%)} \\
\midrule
Full Model                            & \textbf{89.34±0.24} & 89.25±0.37 \\
w/o BM Encoder                        & 88.36±0.13 & 88.29±0.24 \\
w/o Multi-scale Conv                  & 89.17±0.19 & 88.98±0.28 \\
w/o BiLSTM                            & 89.21±0.17 & \textbf{89.27±0.22} \\
w/o BM Encoder + Multi-scale Conv     & 87.91±0.13 & 87.76±0.38 \\
w/o Multi-scale Conv + BiLSTM         & 89.17±0.21 & 89.20±0.24\\
w/o BiLSTM + BM Encoder               & 85.59±0.20 & 85.47±0.18 \\
\bottomrule
\end{tabular}
\end{table}

\section{Conclusion}
\label{sec:conclusion}

In this work, we introduce MEBM-Speech, a multi-scale enhanced decoding framework tailored for MEG-based speech activity detection.
By combining the BM encoder, multi-scale convolutional branch, and BiLSTM with a depthwise separable fusion layer, the model effectively integrates complementary temporal dynamics across multiple scales.
Our probabilistic decoding strategy with temporal jittering and pooling enhances robustness to onset misalignment and label noise.
The resulting system achieves approximately 89\% F1\textsubscript{macro} on both local and online benchmarks, demonstrating strong generalization across sessions.

Future work will focus on extending the framework toward real-time speech decoding, exploring cross-subject adaptation for broader generalization, and exploring MEG-based speech detection during speech production, moving closer to practical speech–BCI applications for communication restoration.

\begin{ack}
This work was supported by the STI 2030—Major Projects (No. 2021ZD0201500), the High-performance Computing Platform of Peking University, and the Biomedical Computing Platform of National Biomedical Imaging Center of Peking University. We also gratefully acknowledge the guidance and valuable discussions provided by Prof. Jing Chen.
\end{ack}


\appendix


\end{document}